\documentclass[11pt]{article}

 \csname
@addtoreset\endcsname{equation}{section}

\usepackage{latexsym}
\usepackage{amsmath}
\usepackage{mathrsfs}
\def\prd{\pr \cdot}

\def\gz0{\gamma^{0}}

\def\nn{\nonumber}

\def\a{\alpha}
\def\b{\beta}
\def\g{\gamma}

\def\d{\delta}

\def\e{\epsilon}

\def\h{\eta}

\def\l{\lambda}
\def\L{\Lambda}
\def\m{\mu}
\def\n{\nu}
\def\x{\xi}

\def\r{\rho}

\def\vf{\varphi}
\def\c{\chi}

\def\O{\Omega}

\def\cA{{\cal A}}
\def\cB{{\cal B}}
\def\cC{{\cal C}}

\def\cE{{\cal E}}
\def\cF{{\cal F}}

\def\cL{{\cal L}}

\def\cR{{\cal R}}
\def\cS{{\cal S}}

\def\cW{{\cal W}}

\def\cY{{\cal Y}}
\def\cZ{{\cal Z}}

\def\be{\begin{equation}}
\def\ee{\end{equation}}
\def\bea{\begin{eqnarray}}
\def\eea{\end{eqnarray}}
\def\ba{\begin{array}}
\def\ea{\end{array}}
\def\bec{\begin{center}}
\def\ec{\end{center}}
\def\ba{\begin{align}}
\def\ena{\end{align}}

\def\12{\frac{1}{2}}

\def\pr{\partial}
\def\prd{\partial \cdot}
\def\bra{\langle \,}
\def\ket{\, \rangle}
\def\comma{\,,\,}

\def\dsl{\not {\! \pr}}

\def\psisl{\not {\!\! \psi}}

\def\esl{\not {\! \epsilon}}

\def\ssl{\not {\! \cal S}}

\def\ssl{\not {\! \cal S}}

\begin{document}

\begin{center}

%%%%%%%%%%%%%%%%%%%%%%%%%%%%%%%%%%%%%%%%%%%%%%%%%%%%%%%%%%%%%%%%%%%%

{\Large\bfseries Lagrangian formulations for Bose and Fermi\\[5pt] higher-spin fields of mixed symmetry}\\

%%%%%%%%%%%%%%%%%%%%%%%%%%%%%%%%%%%%%%%%%%%%%%%%%%%%%%%%%%%%%%%%%%%%

\vspace{25pt}
{\sc A.~Campoleoni${}^{\; a,}$\footnote{Address after October 1, 2009: Max-Planck-Institut f\"ur Gravitationsphysik, Albert-Einstein-Institut,
Am M\"uhlenberg 1, DE-14476 Golm, Germany.}}\\[15pt]

{${}^a$\sl\small
Scuola Normale Superiore and INFN\\
Piazza dei Cavalieri, 7\\I-56126 Pisa \ ITALY \\
e-mail: {\small \it a.campoleoni@sns.it}}

\vspace{20pt}

\end{center}

\begin{abstract}
We review the structure of local Lagrangians and field equations for free bosonic and fermionic gauge fields of mixed symmetry in flat space. These are first presented in a constrained setting and then the ($\g$-)trace constraints on fields and gauge parameters are eliminated via the introduction of a number of auxiliary fields.

\vspace{25pt}
\noindent
\emph{Based on the talk presented at the XVIII SIGRAV Conference on General Relativity and Gravitational Physics, Cosenza, 22-25 September 2008}
\end{abstract}
\vspace{2pt}

%%%%%%%%%%%%%%%%%%%%%%%%%%%%%%%%%%%%%%%%%%%%%%%

\section{Introduction}\label{sec:intro}

%%%%%%%%%%%%%%%%%%%%%%%%%%%%%%%%%%%%%%%%%%%%%%%

The current understanding of higher-spin gauge theories is incomplete mainly because the systematics of their interactions is still rather obscure\footnote{For reviews of the current status of higher-spin gauge theories see \cite{review} and the references therein.}. However, even a number of issues concerning the free theory were clarified only recently, and this work reviews those discussed in \cite{mixed_bose,mixed_fermi} and regarding fields of mixed symmetry. These types of fields are necessary to describe all the irreps of the Poincar\'e group whenever the space-time dimension is greater than five, and are key components of the massive spectra of String Theory. On the other hand, the $\a^{\,\prime}$ dependence of couplings and masses has long suggested the widely held idea that String Theory could well be a broken phase of a higher-spin gauge theory\footnote{See for instance \cite{string}, that also contains some references to the original literature.}, although the breaking mechanism at work still lacks a precise formulation. The quest for proper tools to analyze these problems is indeed one of the main motivations behind much of the current literature on higher spins, and in particular of \cite{mixed_bose,mixed_fermi}, where different Lagrangian formulations for massless mixed-symmetry fields in flat space are built along the lines of the metric formalism for gravity. At any rate, massless and massive models are strictly related, since the latter can be obtained from the former via a Kaluza-Klein circle reduction.

Actions for free gauge fields with spin $s>2$ were first constructed by Fronsdal and Fang in the late seventies \cite{fronsdal} considering the massless limit of the massive Singh-Hagen construction \cite{singhagen}, that also displayed the expected emergence of a gauge symmetry. The fields considered in these works are fully symmetric tensors $\vf_{\,\m_1 \ldots\, \m_s}$ and spinor-tensors $\psi^{\,\a}{}_{\m_1 \ldots\, \m_s}$, that generalize the vector potential $A_{\,\m}$ and the linearized metric fluctuation $h_{\,\m\n}$. In partial analogy with these ``low-spin'' examples, they are supplemented by the gauge transformations
\begin{align}
& \d\, \vf_{\,\m_1 \ldots\, \m_s} \, = \, \pr_{\,(\,\m_1\,} \L_{\,\m_2 \ldots\, \m_s\,)} \, , \nn \\
& \d\, \psi^{\,\a}{}_{\m_1 \ldots\, \m_s} \, = \, \pr_{\,(\,\m_1\,} \e^{\,\a}{}_{\m_2 \ldots\, \m_s\,)} \, , \label{gauge_symm}
\end{align}
where here and in the following a couple of parentheses denotes a complete symmetrization of the indices it encloses, with the minimum possible number of terms and with unit overall normalization. However, in the Fang-Fronsdal formulation \emph{fields and gauge parameters are to satisfy some algebraic ($\g$-)trace constraints}. The constraints on the gauge parameters can be conveniently identified looking at the non-Lagrangian field equations, that for generic spin $s$ or $s+\12$ fields read
\begin{align}
& \cF_{\,\m_1 \ldots\, \m_s} \, \equiv \ \Box \, \vf_{\,\m_1 \ldots\, \m_s} \, - \, \pr_{\,(\,\m_1\,} \prd \vf_{\,\m_2 \ldots\, \m_s\,)} \, + \, \pr_{\,(\,\m_1\,}\pr_{\phantom{(}\!\m_2}\, \vf_{\,\m_3 \ldots\, \m_s\,)\,\l}{}^{\,\l} \, = \, 0 \, , \nn \\[5pt]
& \cS_{\,\m_1 \ldots\, \m_s} \, \equiv \ i \, \left\{\, \dsl \ \psi_{\,\m_1 \ldots\, \m_s} \, - \, \pr_{\,(\,\m_1} \psisl{}_{\,\m_2 \ldots\, \m_s\,)} \,\right\} \, = \, 0 \, , \label{fronsdal}
\end{align}
where we hid the spinor index carried by $\psi$, as we shall do in the following for all spinor indices.
Indeed, while in the spin $2$ case the equation of motion
\be \label{ricci}
R_{\,\m\n} \, \equiv \ \Box \, h_{\,\m\n} \, - \, \pr_{\,\m}\, \prd h_{\,\n} \, - \, \pr_{\,\n}\, \prd h_{\,\m} \, + \, \pr_{\,\m\,} \pr_{\,\n} \, h_{\,\l}{}^{\,\l} \, = \, 0
\ee
sets to zero the linearized Ricci tensor that provides a gauge invariant completion of $\Box\, h_{\,\m\n}$, for $s > 2$ eqs.~\eqref{fronsdal} are invariant under the gauge transformations \eqref{gauge_symm} if and only if one imposes the algebraic constraints
\be \label{constrg_symm}
\L_{\,\m_1 \ldots\, \m_{s-3}\,\l}{}^{\,\l} \, = \, 0 \, , \qquad\quad\qquad \esl_{\,\m_1 \ldots \m_{s-2}} \, = \, 0 \, .
\ee
The constraints on the gauge fields can be conveniently identified looking at the generalizations of the Bianchi identity of linearized gravity,
\be \label{bianchi_grav}
\prd R_{\,\m} \, - \, \12 \ \pr_{\,\m}\, R_{\,\l}{}^{\,\l} \, = \, 0 \, ,
\ee
that read
\begin{align}
& \prd \cF_{\,\m_1 \ldots\, \m_{s-1}} \, - \, \12 \ \pr_{\,(\,\m_1}\, \cF_{\,\m_2 \ldots\, \m_{s-1}\,)\,\l}{}^{\,\l} \, = \, - \, \frac{1}{4}\ \pr_{\,(\,\m_1\,}\pr_{\phantom{(}\!\m_2\,}\pr_{\phantom{(}\!\m_3}\, \vf_{\,\m_4 \ldots\, \m_{s-1}\,)\,\l\r}{}^{\,\l\r} \, , \label{bianchi_symm} \\[5pt]
& \prd \cS_{\,\m_1 \ldots\, \m_{s-1}} \, - \, \12 \dsl \ssl_{\,\m_1 \ldots\, \m_{s-1}} - \, \12 \ \pr_{\,(\,\m_1}\, \cS_{\,\m_2 \ldots\, \m_{s-1}\,)\,\l}{}^{\,\l} = \, \frac{i}{2} \ \pr_{\,(\,\m_1\,}\pr_{\phantom{(}\!\m_2} \psisl{}_{\,\m_3 \ldots\, \m_{s-1}\,)\,\l}{}^{\,\l} \, . \nn
\end{align}
In fact, the Fang-Fronsdal constraints on the fields are
\be \label{constrf_symm}
\vf_{\,\m_1 \ldots\, \m_{s-4}\,\l\r}{}^{\,\l\r} \, = \, 0 \, , \qquad\qquad \psisl{}_{\,\m_1 \ldots\, \m_{s-3}\,\l}{}^{\,\l} \, = \, 0 \, ,
\ee
and eliminate the right-hand sides of the Bianchi identities. Their role can be understood recalling that the Bianchi identity \eqref{bianchi_grav} determines the structure of the divergenceless linearized Einstein tensor. In strict analogy, gauge invariant Lagrangians for higher-spin fields can be recovered enforcing the constraints \eqref{constrf_symm}.

As we anticipated, fully symmetric fields do not suffice to describe all irreps of the Poincar\'e group in generic space-time dimensions. One is thus led to consider their ``mixed-symmetry'' generalizations $\vf_{\,\m_1 \ldots\, \m_{s_1};\, \n_1 \ldots\, \n_{s_2}; \,\ldots}$ and $\psi^{\,\alpha}{}_{\m_1 \ldots\, \m_{s_1};\, \n_1 \ldots\, \n_{s_2}; \,\ldots}$, possessing several groups of symmetrized space-time indices\footnote{Without imposing any symmetry relating the various sets of indices, these fields can only describe reducible representations of the Poincar\'e group. However, $gl(D)$ reducible (spinor-)tensors naturally emerge in String Theory, where they are associated with products of bosonic oscillators. It is thus natural and convenient to work with them. Furthermore, the Young projectors extracting irreducible $gl(D)$ components from a given (spinor-) tensor commute with field equations and Lagrangians, so that the relevant information for irreducible fields can be easily extracted from those presented in the following. For further details on this issue we refer the reader to \cite{mixed_bose,mixed_fermi}, where we also discussed how to adapt the formalism to multi-forms possessing several groups of antisymmetrized space-time indices.}. The development of String Field Theory stimulated the study of these fields in the eighties \cite{old_mixed}, and finally Labastida proposed the gauge transformations, the equations of motion and the set of constraints forcing arbitrary Bose and Fermi fields of mixed symmetry to propagate free massless modes \cite{laba_eq}. Furthermore, in \cite{laba_lag} he also identified the Lagrangians leading to the equations of motion for constrained Bose fields, while their fermionic counterparts were obtained only recently in \cite{mixed_fermi}\footnote{Actually, a covariant Lagrangian formulation for arbitrary irreps of the Poincar\'e group was first obtained in \cite{siegel}, but within a BRST-like setup that is rather remote from these developments. Another alternative formulation of higher-spin dynamics was recently presented in \cite{staunton}.}.

In order to deal with mixed-symmetry fields it is convenient to resort to the compact notation of \cite{mixed_bose,mixed_fermi}, where all space-time indices are hidden. On the other hand, \emph{``family'' indices} (in the following denoted by small-case latin letters) select the groups of space-time indices that gradients $\pr^{\,i}$, divergences $\pr_{\,i}$ and traces $T_{ij}$ are acting upon\footnote{A further useful convention of \cite{mixed_bose,mixed_fermi} states that \emph{lower} family indices are associated to operators removing Lorentz indices, while \emph{upper} family indices are associated to operators adding Lorentz indices, to be symmetrized with all the other indices of the group identified by the given family label. As in the rest of the paper, the symmetrizations involve the minimum possible number of terms and unit overall normalization, while the Einstein convention for summing over pairs of indices is used throughout. At any rate, \emph{all the needed information is contained in the algebra satisfied by the basic operators}, and in the present section the relevant rules are
\begin{equation*}
[\, \pr_{\,i} \comma \pr^{\,j} \,] \, = \, \Box \, \d_{\,i}{}^{\,j} \, , \qquad
[\, \g_{\,i} \comma \pr^{\,j} \,] \, = \ \dsl \, \d_{\,i}{}^{\,j} \, , \qquad
[\, T_{ij} \comma \pr^{\,k} \,] \, = \, \pr_{\,(\,i}\, \d_{\,j\,)}{}^{\,k} \, .
\end{equation*}}.
In a similar fashion, family indices are associated to $\g$-traces $\g_{\,i}$, and it also proves convenient to introduce their antisymmetric combinations like $\g_{\,ij} = \12 \left(\, \g_{\,i}\,\g_{\,j} - \g_{\,j}\,\g_{\,i} \,\right)$. In this notation whole classes of mixed-symmetry fields can be treated at the same time, since all the resulting expressions only depend on the number of index ``families'' involved, and not on the number of space-time indices they contain. Thus, the Labastida gauge transformations, with one independent parameter for each index family, read
\be \label{gauge}
\d \, \vf \, = \, \pr^{\,i}\, \L_{\,i} \, , \qquad\qquad\qquad \d \, \psi \, = \, \pr^{\,i} \, \e_{\,i} \, ,
\ee
where the parameter labeled by the index $i$ carries only $(s_i-1)$ indices in the $i$-th group, to be symmetrized with the one carried by the gradient. Furthermore, the Labastida equations of motion take the compact form
\begin{align}
& \cF  \, \equiv \, \Box \, \vf \, - \, \pr^{\,i}\,\pr_{\,i} \, \vf \, + \, \12 \ \pr^{\,i}\pr^{\,j}\, T_{ij}\, \vf \, = \, 0 \, , \nn \\[5pt]
& \cS \, \equiv \, i \, \left\{\, \dsl \, \psi \, - \, \pr^{\,i} \, \g_{\,i}\, \psi \,\right\} \, = \, 0 \, , \label{equations}
\end{align}
where, for instance, the first line of \eqref{equations} is a shorthand for
\begin{align}
& \cF_{\,\m_1 \ldots\, \m_{s_1};\, \n_1 \ldots\, \n_{s_2}; \,\ldots} \equiv \, \Box \, \vf_{\,\m_1 \ldots\, \m_{s_1};\, \n_1 \ldots\, \n_{s_2}; \,\ldots} - \, \pr_{\,(\,\m_1|}\, \pr^{\,\l}\, \vf_{\,|\,\m_2 \ldots\, \m_{s_1}\,)\,\l\,;\, \n_1 \ldots\, \n_{s_2}; \,\ldots} \quad \nn \\
& - \, \pr_{\,(\,\n_1|}\, \pr^{\,\l}\, \vf_{\,\m_1 \ldots\, \m_{s_1};\, |\,\n_2 \ldots\, \n_{s_2}\,)\,\l\,; \,\ldots} - \, \ldots \, + \, \pr_{\,(\,\n_1\,|\,} \pr_{\,(\,\m_1}\, \vf_{\,\m_2 \ldots\, \m_{s_1}\,)\,\l\,;\,|\,\n_2 \ldots\, \n_{s_2}\,)}{}^{\l}{}_{\,;\,\ldots} \nn \\
& + \, \pr_{\,(\,\m_1} \pr_{\phantom{(}\!\m_2}\, \vf_{\,\m_3 \ldots\, \m_{s_1}\,)\,\l}{}^{\l}{}_{\,;\,\n_1 \ldots\, \n_{s_2};\,\ldots} + \, \pr_{\,(\,\n_1} \pr_{\,\n_2\,|}\, \vf_{\,\m_1 \ldots\, \m_{s_1};\,|\,\n_3 \ldots\, \n_{s_2}\,)\,\l}{}^{\l}{}_{\,;\,\ldots} + \, \ldots \, \, = \, 0 \, , \label{short}
\end{align}
in which the omitted terms involve the remaining index families in a similar fashion.
Eqs.~\eqref{equations} are gauge invariant if and only if
\be \label{constrg}
T_{(\,ij}\, \L_{\,k\,)} \, = \, 0 \, , \qquad\qquad\qquad \g_{\,(\,i}\, \e_{\,j\,)} \, = \, 0 \, ,
\ee
so that, differently from the Fang-Fronsdal symmetric setting, now \emph{not} all ($\g$-)traces of the gauge parameters have to vanish. The same difference emerges looking at the Bianchi identities: for mixed-symmetry gauge fields they read
\begin{align}
& \pr_{\,i}\, \cF \, - \, \12 \ \pr^{\,j}\, T_{ij} \, \cF \, = \, - \, \frac{1}{12} \ \pr^{\,j}\pr^{\,k}\pr^{\,l} \, T_{(\,ij}\,T_{kl\,)} \, \vf \, , \nn \\[5pt]
& \pr_{\,i}\, \cS \, - \, \12 \dsl \, \g_{\,i}\, \cS \, - \, \12 \ \pr^{\,j}\,T_{ij}\, \cS \, - \, \frac{1}{6} \ \pr^{\,j}\,\g_{\,ij}\,\cS \, = \, \frac{i}{6} \ \pr^{\,j}\pr^{\,k}\, T_{(\,ij}\,\g_{\,k\,)}\, \psi \, , \label{bianchi}
\end{align}
and their right-hand sides only contain special linear combinations of ($\g$-)traces acting on the fields. These identify the Labastida constraints
\be \label{constrf}
T_{(\,ij}\,T_{kl\,)}\, \vf \, = \, 0 \, , \qquad\qquad\qquad T_{(\,ij}\,\g_{\,k\,)}\, \psi \, = \, 0 \, ,
\ee
that one has to force upon the fields in order to build gauge invariant Lagrangians. Indeed, as we shall review in Section \ref{sec:constrained}, the Bianchi identities provide a neat rationale leading to the Lagrangians of \cite{laba_lag,mixed_bose,mixed_fermi}.

Having presented the basic elements of the \emph{constrained} formulation for arbitrary higher-spin fields, we should stress that in String Field Theory \emph{no} constraints are present. Indeed, starting from the late nineties, a number of papers have proposed alternative formulations of the free dynamics of single higher-spin modes without any need for constraints. These works comprise two main groups, since the constraints can be eliminated both via the introduction of non-local terms and via the introduction of auxiliary fields. The first approach was developed in \cite{symm_nl,current,dario} for fully symmetric Bose and Fermi gauge fields and was then extended to mixed-symmetry fields in \cite{mixed_nl}. It has the virtue of leading to a geometric description, where Lagrangians and equations of motion are built from the linearized curvatures introduced by de Wit and Freedman \cite{dewfr}, or from their mixed-symmetry generalizations \cite{mixed_nl}. On the other hand, the second approach \cite{bpt,symm_l,current,triplets,dario} leads to more standard local Lagrangians, and in Section \ref{sec:unconstrained} the ``minimal'' formulation for fully symmetric fields \cite{symm_l,current} will be extended to arbitrary mixed-symmetry fields following \cite{mixed_bose,mixed_fermi}. In this setup the number of auxiliary fields only depends on the number of index families, and not on the total number of space-time indices as in previous works \cite{bpt}. Other descriptions of the free dynamics of mixed-symmetry fields in flat space or in A(dS) were also built along the lines of the frame-like formalism for gravity \cite{frame}, in a light-cone formalism \cite{light}, or resorting to BRST techniques \cite{mixed_brst}. Furthermore, some steps toward a metric-like description of mixed-symmetry fields in constant curvature spaces were performed in \cite{ads}.

%%%%%%%%%%%%%%%%%%%%%%%%%%%%%%%%%%%%%%%%%%%%%%%

\section{Constrained Lagrangians for Bose and Fermi fields}\label{sec:constrained}

%%%%%%%%%%%%%%%%%%%%%%%%%%%%%%%%%%%%%%%%%%%%%%%

As it is well known, the field equation \eqref{ricci} is non-Lagrangian, but is in general equivalent to that following from the linearized Einstein-Hilbert Lagrangian
\be \label{lag_grav}
\cL \, = \, \12 \ h^{\,\m\n} \left(\, R_{\,\m\n} \, - \, \12 \ \h_{\,\m\n}\, R_{\,\l}{}^{\,\l} \,\right) \, .
\ee
The Lagrangian \eqref{lag_grav} contains the linearized Ricci tensor appearing in \eqref{ricci} and its only available trace. Furthermore, the relative coefficient entering \eqref{lag_grav} is fixed uniquely by the request of gauge invariance. In fact, up to total derivatives, the gauge variation of \eqref{lag_grav} under the linearized diffeomorphisms $\d\, h_{\,\m\n} = \, \pr_{\,(\,\m}\,\x_{\,\n\,)}$ is
\be \label{var_grav}
\d\, \cL \, = \, - \, \12 \ \x^{\,\m} \left(\, \prd R_{\,\m} \, - \, \12 \ \pr_{\,\m}\, R_{\,\l}{}^{\,\l} \,\right) \, = \, 0 \, ,
\ee
that vanishes on account of the Bianchi identity \eqref{bianchi_grav}. The equations of motion \eqref{fronsdal} for symmetric bosons are non-Lagrangian as well, but the Fronsdal constraint \eqref{constrf_symm} on $\vf$ also implies the vanishing of the double trace of the tensor $\cF$ appearing in \eqref{fronsdal}. Thus, Lagrangians leading to equations equivalent to \eqref{fronsdal} can only contain $\cF$ and its first trace, again with a relative coefficient that is fixed by the Bianchi identities via the requirement of gauge invariance, so that the result takes a form similar to \eqref{lag_grav}. Actually, even symmetric fermions can be treated in this fashion, since the triple $\g$-trace of $\cS$ is forced to vanish by the constraint \eqref{constrf_symm} on $\psi$.

In the mixed-symmetry case the tensors $\cF$ and $\cS$ of \eqref{equations} satisfy
\be \label{constrF}
T_{(\,ij}\,T_{kl\,)} \, \cF \, = \, 0 \, , \qquad\qquad T_{(\,ij}\,\g_{\,k\,)} \, \cS \, = \, 0 \, ,
\ee
so that the Labastida constraints \eqref{constrf} on the fields induce similar constraints on the kinetic (spinor-)tensors. However, differently from the symmetric setup, in this case \emph{not} all double traces of $\cF$ and \emph{not} all triple $\g$-traces of $\cS$ vanish. The Lagrangians thus can, and indeed do, contain combinations of multiple ($\g$-)traces of the kinetic (spinor-)tensors. These can be identified decomposing the available expressions according to the irreps of the \emph{permutation group acting on the family indices} carried by the operators $T_{ij}$ or $\g_{\,i}$. Actually, eqs.~\eqref{constrF} imply that only \emph{two-column} Young projected combinations can enter the Lagrangians\footnote{Projections associated with Young diagrams with at least three columns are realized via a sum of terms with at least three symmetrized indices. These can be always manipulated to rebuild the constraints \eqref{constrF}. Further details on this statement and on the needed tools related to the symmetric group can be found in the appendices of \cite{mixed_bose,mixed_fermi}.}, for both Bose and Fermi fields. Clearly, they have to be contracted with suitable products of invariant tensors, so that the natural ansatz for the bosonic Lagrangians \cite{laba_lag,mixed_bose} reads
\be \label{lag_bose}
\cL \, = \, \12 \ \bra\, \vf \,\comma\, \sum_{p\,=\,0}^N \, k_{\,p}\ \h^{i_1j_1} \ldots\, \h^{i_pj_p}\, Y_{\{2^p\}}\, T_{i_1j_1} \ldots\, T_{i_pj_p} \, \cF \,\ket \, ,
\ee
since a product of identical $T_{ij}$ tensors only admits projections associated to Young diagrams with an even number of boxes in each row. As a result, the only available two-column projection for a product of $p$ traces is the $\{2^p\}$ one, corresponding to a rectangular diagram with two columns and $p$ rows. Furthermore, in \eqref{lag_bose} $N$ denotes the number of index families and we have introduced a convenient scalar product\footnote{The scalar product is defined as
\begin{equation*}
\bra \vf \comma \c \ket \, \equiv \, \frac{1}{s_1! \ldots s_n!} \ \vf_{\m^1_1 \ldots \, \m^1_{s_1} \, ; \, \ldots \, ; \,
\m^n_1 \ldots \, \m^n_{s_n}} \, \c^{\,\m^1_1 \ldots \, \m^1_{s_1} \, ; \, \ldots \, ; \, \, \m^n_1 \ldots \, \m^n_{s_n}} \, \equiv \,
\frac{1}{s_1! \, \ldots s_n!} \ \vf \, \c \, ,
\end{equation*}
and allows to integrate by parts without introducing $s_i$-dependent combinatoric factors. Notice that, in order to recover the usual normalizations, an overall factor $\prod_{i=1}^N s_i!$ should accompany the Lagrangians. However, for brevity this factor is ignored in all the expressions appearing in this paper.} following \cite{mixed_bose,mixed_fermi}. The definition of the $\h^{\,ij}$ operators entering \eqref{lag_bose} is quite natural,
\be \label{metric}
\h^{\,ij} \, \vf \, \equiv \, \12 \, \sum_{n\,=\,1}^{s_i+1}  \, \h_{\,\m^i_n \, (\,\m^j_1\,|}\ \vf_{\,\ldots \,;\, \ldots\, \m^i_{r\,\neq\,n} \ldots\, ;\, \ldots
\,;\, |\,\m^j_2\, \ldots\, \m^j_{s_j+1}) \,;\, \ldots} \ ,
\ee
aside from the unconventional factor $1/2$, that proves convenient in the presentation of a number of results. In a similar fashion, for Fermi fields the ansatz for the structure of the Lagrangians \cite{mixed_fermi} is
\be \label{lag_fermi}
\cL  = \, \12 \, \bra\, \bar{\psi} \comma \sum_{p\,,\,q\,=\,0}^N  k_{\,p\,,\,q}\ \h^{i_1j_1} \ldots\, \h^{i_pj_p}\, \g^{\,k_1 \ldots\, k_q}\, Y_{\{2^p,1^q\}}\, T_{i_1j_1} \ldots\, T_{i_pj_p} \g_{\,k_1 \ldots\, k_q} \, \cS \,\ket \, + \, \textrm{h.c.} \, .
\ee
The multiple $\g$-traces of $\cS$ are here written in an antisymmetric basis with
\be \label{gamma}
\g_{\,k_1 \ldots\, k_q} \, = \, \frac{1}{q\,!} \ \g_{\,[\,k_1} \g_{\phantom{[}\!k_2} \ldots\, \g_{\,k_q\,]} \, ,
\ee
where the square brackets denote the antisymmetrization of the indices they enclose, again with overall factor one. The coefficients $k_{\,p}$ and $k_{\,p\,,\,q}$ appearing in \eqref{lag_bose} and \eqref{lag_fermi} can be fixed requiring the \emph{gauge invariance} of the Lagrangians, in strict analogy with the Einstein-Hilbert case \eqref{lag_grav}.

The gauge variation of the bosonic Lagrangian \eqref{lag_bose} reads
\begin{align}
\d \, \cL \, = \, & - \, \sum_{p\,=\,0}^N \, \frac{1}{2^{\,p+1}} \, \bra\, Y_{\{2^p,1\}}\, T_{i_1j_1} \!\ldots T_{i_pj_p}\, \L_{\,k} \,\comma\, k_{\,p}\ Y_{\{2^p,1\}}\, \pr_{\,k}\, T_{i_1j_1} \!\ldots T_{i_pj_p} \, \cF  \nn \\
& + \, (\,p+1\,)\, k_{\,p+1}\ \pr^{\,l}\, Y_{\{2^{p+1}\}}\, T_{i_1j_1} \ldots\, T_{i_pj_p} T_{kl}\, \cF \ket \, . \label{lag_var}
\end{align}
Indeed, the other available Young projections of the divergence terms, the $\{3,2^{p-1}\}$ ones, would reconstruct the constraints \eqref{constrg} on the gauge parameters on the left entry of the scalar products. On the other hand, computing the $\{2^p,1\}$ Young projection in the family indices carried by a product of  $p$ \emph{traces of the Bianchi identities} \eqref{bianchi}, gives
\be \label{trace_bianchi}
(\,p+2\,)\ Y_{\{2^p,1\}}\, \pr_{\,k}\, T_{i_1j_1} \ldots\, T_{i_pj_p} \, \cF \, - \, \pr^{\,l}\, Y_{\{2^{p+1}\}}\, T_{i_1j_1} \ldots\, T_{i_pj_p} T_{kl}\, \cF \, = \, 0 \, ,
\ee
up to the constraints \eqref{constrf} that we enforced on the right-hand side of \eqref{bianchi}. One can then rebuild the combinations \eqref{trace_bianchi} in \eqref{lag_var} selecting the values
\be \label{coeff_bose}
k_{\,p} \, =  \, \frac{(-1)^{\,p}}{p\,!\,(\,p+1\,)\,!} \ ,
\ee
that lead to a gauge invariant result. These are the coefficients first obtained by Labastida in \cite{laba_lag}, barring a slight change of notation due to the definition \eqref{metric} and a typo in the oscillating sign that was corrected in \cite{mixed_bose}.

For brevity, we refrain from repeating this procedure also for Fermi fields since a detailed discussion can be found in \cite{mixed_fermi}. Due to the presence of $\g$-traces this involves some technical complications, but the logical steps are essentially those followed for Bose fields and already emerging in the simpler case of linearized gravity. One can fix the coefficients $k_{\,p\,,\,q}$ in the Lagrangian \eqref{lag_fermi} computing its gauge variation and comparing the result with the two-column Young projected $\g$-traces of the Bianchi identities \eqref{bianchi}. In order to do that, one must first eliminate from the gauge variation the terms proportional to the constraints \eqref{constrg} on the gauge parameters. The remainder is then two-column projected, and with a proper choice of the $k_{\,p\,,\,q}$ reproduces the two-column projected consequences of the Bianchi identities. This procedure fixes uniquely the coefficients and the result, first presented in \cite{mixed_fermi}, reads
\be \label{coeff_fermi}
k_{\,p\,,\,q} \, =  \, \frac{(-1)^{\,p\,+\,\frac{q\,(q+1)}{2}}}{p\,!\,q\,!\,(\,p+q+1\,)\,!} \ .
\ee

One can also obtain the coefficients \eqref{coeff_bose} and \eqref{coeff_fermi} looking for self-adjoint Einstein-like tensors \cite{mixed_fermi}, following the approach of \cite{laba_lag}. In fact, since $\cF$ and $\cS$ are gauge invariant in the constrained theory, building their self-adjoint extensions in \eqref{lag_bose} and \eqref{lag_fermi} leads to a gauge invariant result. As the reader can expect, the constraints \eqref{constrf} on the fields that affect the Bianchi identities \eqref{bianchi} are crucial also in this approach.

The Lagrangians \eqref{lag_bose} take a particularly simple form for $gl(D)$-\emph{irreducible} two-column Bose fields. They are characterized by an arbitrary number of index families containing at most two space-time indices and by the conditions that all the symmetrizations over any group of three indices vanish. This form obtains expressing \eqref{lag_bose} directly in terms of the fields, and generalizing a similar rewriting of the linearized Einstein-Hilbert Lagrangian. Indeed, up to total derivatives, the Lagrangian \eqref{lag_grav} can be cast in the form
\be \label{antigrav}
\cL \, = \, - \, \12 \ \pr^{\,\m}\, h^{\,\n}{}_{\,[\,\m}\, \pr_{\phantom{[}\!\n}\, h_{\,\r\,]}{}^{\,\r} \, .
\ee
In a similar fashion \cite{rewriting,mixed_bose}, for $gl(D)$-irreducible two-column Bose fields of the form $\{2^p,1^q\}$ the Lagrangian \eqref{lag_bose} has the structure
\begin{align}
\cL \, \sim & \ \pr^{\,\m_1} \, \vf^{\,\m_2}{}_{\,[\,\m_1\,|\,; \,\ldots\, ;}{}^{\,\m_{p+1}}{}_{\,|\,\m_p;\,\m_{p+1}; \,\ldots\, ;\,\m_{p+q}\,|} \nn \\
& \, \times \, \pr_{\,|\,\m_{p+q+1}}\, \vf_{\,\m_{p+q+2}\,|}{}^{\,\m_{p+2}}{}_{\,; \,\ldots\, ;\,|\,\m_{2p+q+1}\,]}{}^{\,\m_{2p+1}; \,\m_{2p+2}; \,\ldots\, ;\,\m_{2p+q+1}} \, . \label{antilagb}
\end{align}
While the direct comparison of the two results \eqref{lag_bose} and \eqref{antilagb} is not straightforward, the latter Lagrangian has to coincide with \eqref{lag_bose} up to an overall coefficient, since it is \emph{manifestly} gauge invariant. In fact, the gradients coming from the gauge transformations are always antisymmetrized with one of the derivatives already present in the Lagrangian. Notice furthermore that the fields in this class are fully \emph{unconstrained}, since the Labastida constraints \eqref{constrg} and \eqref{constrf} are not available due to the irreducibility condition. However, eq.~\eqref{antilagb} fixes the coefficients \eqref{coeff_bose}, that already show up in the Lagrangians of two-column fields. Thus, labeling arbitrary irreps of the Poincar\'e group with Young diagrams, it is clear that the need for constraints is dictated by the number of columns, while the appearance of higher-trace terms in the Lagrangians is dictated by the number of rows.
Similar considerations apply to Fermi fields, where the usual presentation of the Rarita-Schwinger Lagrangian,
\be \label{antirarita}
\cL \, = \, \frac{i}{2} \ \psi_{\,\m}\, \g^{\,\m\n\r}\, \pr_{\,\n}\, \psi_{\,\r} \, + \, \textrm{h.c.} \ ,
\ee
extends to arbitrary one-column fully-antisymmetric fields via \cite{mixed_fermi}
\be \label{antilagf}
\cL \, = \, i \ \frac{(-1)}{2\,q\,!}^{\frac{q\,(q-1)}{2}} \ \bar{\psi}_{\,\m_1 \ldots\, \m_q}\, \g^{\,\m_1 \ldots\, \m_q  \l \, \n_1 \ldots\, \n_q} \, \pr_{\,\l}\, \psi_{\,\n_1 \ldots\, \n_q}  + \, \textrm{h.c.} \ .
\ee
The Lagrangian \eqref{antilagf} is \emph{manifestly} gauge invariant due to the contraction of all the space-time indices with the fully antisymmetric $\g$-matrix it contains, while one-column Fermi fields are fully \emph{unconstrained}.

%%%%%%%%%%%%%%%%%%%%%%%%%%%%%%%%%%%%%%%%%%%%%%%

\section{Minimal Unconstrained Lagrangians}\label{sec:unconstrained}

%%%%%%%%%%%%%%%%%%%%%%%%%%%%%%%%%%%%%%%%%%%%%%%

In the Lagrangian theory, the Labastida constraints \eqref{constrg} and \eqref{constrf} can be eliminated in a ``minimal'' way in two steps:
\begin{itemize}
\item the constraints \eqref{constrg} on the gauge parameters can be eliminated via the introduction of at most one compensator field for each constraint;
\item the constraints \eqref{constrf} on the fields can be eliminated via the introduction of at most one Lagrange multiplier for each constraint.
\end{itemize}
The correspondence between auxiliary fields and constraints is \emph{not} one-to-one since the Labastida constraints take a very compact form in this notation but are \emph{not} independent: their higher ($\g$-)traces can indeed become proportional. Anyway, following this path one obtains a fully unconstrained Lagrangian formulation, adding a number of auxiliary fields that depend only on the number of index families. The introduction of compensator fields leads directly to equations of motion with unconstrained gauge invariance and was first presented for symmetric fields in the first of \cite{symm_l}. This setup was then complemented by Lagrange multipliers and extended off-shell in the third of \cite{symm_l}, but in the present discussion we shall reverse the order of these two steps. Indeed, the intermediate case of a Lagrangian theory for unconstrained fields still only allowing constrained gauge transformations is also of some interest.

Relaxing the constraints \eqref{constrf} on the fields, the Lagrangians \eqref{lag_bose} and \eqref{lag_fermi} are no longer gauge invariant, due to the ``classical anomalies'' appearing on the right-hand sides of the Bianchi identities \eqref{bianchi}, whose effect can be compensated adding to the Lagrangians the terms
\be \label{lagrange}
\widetilde{\cL}_{\,\textrm{Bose}} \, = \, \frac{1}{24} \, \bra\, \b_{\,ijkl} \,\comma\, T_{(\,ij}\, T_{kl\,)}\, \vf \ket \, \, , \qquad\qquad \widetilde{\cL}_{\,\textrm{Fermi}} \, = \, \frac{i}{12} \, \bra\, \bar{\l}_{\,ijk} \,\comma\, T_{(\,ij}\, \g_{\,k\,)}\, \psi \,\ket \, + \, \textrm{h.c.} \, .
\ee
The gauge transformations of the Lagrange multipliers appearing in \eqref{lagrange} can be fixed integrating by parts the terms generated by these ``classical anomalies'' and their ($\g$-)traces, and the reader can find their detailed form in \cite{mixed_bose,mixed_fermi}. Furthermore, the terms \eqref{lagrange} enforce on-shell the Labastida constraints \eqref{constrf}. Hence, the equations of motion following from the resulting unconstrained Lagrangians are simply related to those following from the constrained ones, as we shall see more in detail in the next section.

One can also obtain Lagrangians for unconstrained fields modifying the structures that appear in \eqref{lag_bose} and \eqref{lag_fermi}, and to this end it is convenient to consider explicitly their $\vf$ and $\psi$ dependence. In this way, one can redefine the ($\g$-)traces of $\cF$ and $\cS$ eliminating altogether the terms proportional to the Labastida constraints \eqref{constrf}, which leads by construction to a gauge invariant result. In fact, the resulting expressions
\begin{align}
& \widehat{\cL}_{\,\textrm{Bose}} = \, \12 \ \bra\, \vf \,\comma\, \sum_{p\,=\,0}^N \, k_{\,p}\ \h^{i_1j_1} \ldots\, \h^{i_pj_p}\, \widehat{\cF}^{\,[\,p\,]}{}_{\,i_1j_1, \,\ldots\, ,\,i_pj_p} \,\ket \, , \nn \\
& \widehat{\cL}_{\,\textrm{Fermi}} = \, \12 \ \bra\, \bar{\psi} \,\comma \!\sum_{p\,,\,q\,=\,0}^N  k_{\,p\,,\,q}\ \h^{i_1j_1} \ldots\, \h^{i_pj_p}\, \g^{\,k_1 \ldots\, k_q}\, \widehat{\cS}^{\,[\,p\,,\,q\,]}{}_{\,i_1j_1, \,\ldots\, ,\,i_pj_p;\,k_1 \ldots\, k_q} \,\ket \, + \, \textrm{h.c.} \, , \label{newlag}
\end{align}
with
\begin{align}
& \widehat{\cF}^{\,[\,p\,]}{}_{\,i_1j_1, \,\ldots\, ,\,i_pj_p} = \, Y_{\{2^p\}}\, T_{i_1j_1} \!\ldots T_{i_pj_p}\, \cF \, - \, \12 \ \pr^{\,k}\pr^{\,l}\, Y_{\{4,2^{p-1}\}}\, T_{i_1j_1} \!\ldots T_{i_pj_p} T_{kl} \, \vf \, , \nn \\[4pt]
& \widehat{\cS}^{\,[\,p\,,\,q\,]}{}_{\,i_1j_1, \,\ldots\, ,\,i_pj_p;\,k_1 \ldots\, k_q} = \, Y_{\{2^p,1^q\}}\, T_{i_1j_1} \!\ldots T_{i_pj_p} \g_{\,k_1 \ldots\, k_q}\, \cS \nn \\
& \phantom{\widehat{\cS}^{\,[\,p\,,\,q\,]}{}_{\,i_1j_1, \,\ldots\, ,\,i_pj_p;\,k_1 \ldots\, k_q} =}\, + \, i \ \pr^{\,l}\, Y_{\{3,2^{p-1},1^q\}}\, T_{i_1j_1} \!\ldots T_{i_pj_p} \g_{\,k_1 \ldots\, k_q l}\, \psi \, , \label{Fmod}
\end{align}
are identical to those effectively entering \eqref{lag_bose} and \eqref{lag_fermi} when one enforces the constraints \eqref{constrf}. Furthermore, these Lagrangians can be supplemented by the terms \eqref{lagrange} that impose on-shell the constraints, but now with gauge invariant Lagrange multipliers. The field equations of these different unconstrained formulations can be combined to provide equivalent conditions and, in fact, one can actually relate them via a field redefinition of the Lagrange multipliers.

Notice that in both formulations the Lagrangians possess an extra symmetry under shifts of particular ($\g$-)traces of the Lagrange multipliers. This is a consequence of the lack of independence of the Labastida constraints that we mentioned above. This fact indeed implies that some combinations of ($\g$-)traces of \eqref{constrf} vanish \emph{identically}, so that for instance
\be
Y_{\{5,1\}}\, T_{mn}\, T_{(\,ij}\,T_{kl\,)}\, \vf \, \equiv \, 0 \, ,
\ee
and consequently the $o(D)$-components of the Lagrange multipliers coupling to them in \eqref{lagrange} can be shifted arbitrarily. A more detailed discussion and similar examples for Fermi fields can be found in Section 2 of \cite{mixed_fermi}.

The constraints \eqref{constrg} on the gauge parameters can be relaxed considering the Stueckelberg-like shifts
\be \label{shift}
\vf \, \to \, \vf \, - \, \pr^{\,i}\, \Phi_{\,i} \, , \qquad\qquad \psi \, \to \, \psi \, - \, \pr^{\,i}\, \Psi_{\,i} \, ,
\ee
with compensator fields transforming as
\be \label{comp_gauge}
\d \, \Phi_{\,i} \, = \, \L_{\,i} \, , \qquad\qquad\qquad \d \, \Psi_{\,i} \, = \, \e_{\,i} \, .
\ee
Under the action of \eqref{shift} arbitrary functions of the fields become gauge invariant, and in particular $\cF$ and $\cS$ give rise to the kinetic (spinor-)tensors
\be \label{kinetic}
\cA \, = \, \cF \, - \, \frac{1}{6} \ \pr^{\,i}\pr^{\,j}\pr^{\,k}\, T_{(\,ij}\, \Phi_{\,k\,)} \, , \qquad\qquad \, \cW \, = \, \cS \, + \, \frac{i}{2} \ \pr^{\,i}\pr^{\,j}\, \g_{\,(\,i}\,\Psi_{\,j\,)} \, ,
\ee
that are invariant under \emph{unconstrained} gauge transformations. However, due to the constrained gauge invariance of $\cF$ and $\cS$, the $\Phi_{\,i}$ and the $\Psi_{\,i}$ enter \eqref{kinetic} \emph{only} via their symmetrized ($\g$-) traces. One can thus eliminate these combinations performing a partial gauge fixing that does not affect the Labastida constrained gauge transformations. This is a crucial condition, since the ``constrained'' portion of the gauge transformations is needed to force the fields to propagate the correct massless modes.

For a symmetric Bose field of spin $s$ the only available combination of the form $\frac{1}{3}\, T_{(\,ij}\,\Phi_{\,k\,)}$ actually coincides with the compensator field $\a_{\,\m_1 \ldots\, \m_{s-3}}$ introduced in \cite{symm_l}. Similarly, for a spin $s+\12$ Fermi field the only available combination of the form $\12\, \g_{\,(\,i}\,\Psi_{\,j\,)}$ coincides with the $\xi_{\,\m_1 \ldots\, \m_{s-2}}$ field of \cite{symm_l}. In principle, even in the mixed-symmetry case one could try to proceed introducing one compensator field for each constraint via
\be \label{alpha}
\a_{\,ijk} \, = \, \frac{1}{3} \, T_{(\,ij}\,\Phi_{\,k\,)} \, , \qquad\qquad \xi_{\,ij} \, = \, \12\, \g_{\,(\,i}\,\Psi_{\,j\,)} \, .
\ee
However, the lack of independence of the Labastida constraints \eqref{constrg} on the gauge parameters leads to the emergence of gauge invariant combinations of the ($\g$-)traces of the $\a_{\,ijk}$ and $\xi_{\,ij}$. Thus, one cannot regard them as independent fields, and resorting to the compensators of \eqref{comp_gauge} seems the most convenient alternative. Further details and some examples of gauge-invariant combinations of the compensators are presented in Section 2 of \cite{mixed_fermi}.

Joining them with suitable shifts of the Lagrange multipliers, the transformations \eqref{shift} can be performed directly in the Lagrangians that we presented before. Again, the Labastida constrained gauge invariance allows to cast the results in the form
\begin{align}
& \cL_{\,\textrm{Bose}} \, = \, \12 \, \bra\, \vf \comma \cE \,\ket \, - \, \frac{1}{8} \, \bra\, \a_{\,ijk}\,(\,\Phi\,) \comma \aleph_{\,ijk}\,(\,\a\,) \,\ket \, + \, \frac{1}{8} \, \bra\, \b_{\,ijkl} \comma \cC_{\,ijkl} \,\ket \, , \nn \\
& \cL_{\,\textrm{Fermi}} \, = \, \12 \, \bra\, \bar{\psi} \comma \cR \,\ket \, + \, \frac{1}{4} \, \bra\, \x_{\,ij}\,(\,\bar{\Psi}\,) \comma \Xi_{\,ij}\,(\,\x\,) \,\ket \, + \, \frac{1}{4} \, \bra\, \bar{\l}_{\,ijk} \comma \cZ_{\,ijk} \,\ket \, + \, \textrm{h.c.} \, , \label{lag_unc}
\end{align}
where only symmetric ($\g$-)traces of the compensators appear. The (spinor-)tensors $\cE$ and $\cR$ of \eqref{lag_unc} can be obtained acting with \eqref{shift} on the combinations appearing respectively in \eqref{lag_bose} and \eqref{lag_fermi}. Thus, they contain combinations of all the ($\g$-)traces of $\cA$ and $\cW$ that are not directly related to the gauge invariant constraints
\be
\cC_{\,ijkl} \, = \, \frac{1}{3} \ T_{(\,ij}\,T_{kl\,)} \left(\, \vf \, - \, \pr^{\,m}\, \Phi_{\,m} \,\right) \, , \qquad\quad \cZ_{\,ijk} \, = \, \frac{i}{3} \ T_{(\,ij}\,\g_{\,k\,)} \left(\, \psi \, - \, \pr^{\,m}\, \Psi_{\,m} \,\right) \, ,
\ee
again with the coefficients displayed in \eqref{coeff_bose} and \eqref{coeff_fermi}. On the other hand, $\aleph$ and $\Xi$ contain suitable combinations of the divergences of $\cE$ and $\cR$, whose explicit forms are rather involved and can be found in \cite{mixed_bose,mixed_fermi}.

The unconstrained $\cA$ and $\cW$, and clearly also the Lagrangians \eqref{lag_unc}, contain higher derivative terms involving the compensators. However, we repeatedly pointed out that these terms can be eliminated by a partial gauge fixing, so that they are really harmless. For instance, for symmetric fields the current-current exchanges obtained coupling an external current to \eqref{lag_bose} or \eqref{lag_fermi} and to \eqref{lag_unc} coincide \cite{current}. The same result is expected to hold in the mixed-symmetry case as well, and this is indeed true in the examples analyzed in \cite{mixed_bose,mixed_fermi}. Nevertheless, more conventional unconstrained formulations with the usual number of derivatives, but still with a spin-independent number of auxiliary fields, were obtained for symmetric fields in \cite{triplets,dario}. Furthermore, in \cite{mixed_bose,mixed_fermi} a similar result was obtained for mixed-symmetry fields in a setting similar to that of \cite{dario}, adding few types of extra fields, whose total number only depends on the number $N$ of index families.

%%%%%%%%%%%%%%%%%%%%%%%%%%%%%%%%%%%%%%%%%%%%%%%

\section{Lagrangian field equations and Weyl like-symmetries}\label{sec:equations}

%%%%%%%%%%%%%%%%%%%%%%%%%%%%%%%%%%%%%%%%%%%%%%%

The combinations of ($\g$-)traces of $\cF$ and $\cS$ entering \eqref{lag_bose} and \eqref{lag_fermi} are self-adjoint in the constrained theory. However, they do not satisfy Labastida-like constraints, even if $\cF$ and $\cS$ actually do \cite{mixed_bose,mixed_fermi}. As a consequence, a projection is needed in order to obtain from the Lagrangians \eqref{lag_bose} and \eqref{lag_fermi} field equations satisfying the same constraints as the gauge fields. Its form is rather involved, and it is thus convenient to deal directly with the equivalent field equations following from the Lagrangians for unconstrained fields, where also the Lagrange multiplier terms \eqref{lagrange} appear. For Bose fields they read
\begin{align}
& \sum_{p\,=\,0}^N \, k_{\,p}\ \h^{i_1j_1} \!\ldots \h^{i_pj_p}\, Y_{\{2^p\}}\, T_{i_1j_1} \ldots\, T_{i_pj_p} \, \cF \, + \, \12 \ \h^{ij}\,\h^{kl}\, \cB_{\,ijkl} \, = \, 0 \, , \nn \\[2pt]
& T_{(\,ij}\,T_{kl\,)}\, \vf \, = 0 \quad \Rightarrow \quad T_{(\,ij}\,T_{kl\,)}\, \cF \, = \, 0 \, , \label{eq_bose}
\end{align}
while for Fermi fields they read
\begin{align}
& \sum_{p\,,\,q\,=\,0}^N  k_{\,p\,,\,q}\ \h^{i_1j_1} \!\ldots \h^{i_pj_p}\, \g^{\,k_1 \ldots\, k_q}\, Y_{\{2^p,1^q\}}\, T_{i_1j_1} \ldots\, T_{i_pj_p} \g_{\,k_1 \ldots\, k_q} \, \cS \, - \, \12 \ \h^{ij}\,\g^{\,k}\, \cY_{\,ijk} \, = \, 0 \, , \nn \\[2pt]
& T_{(\,ij}\,\g_{\,k\,)}\, \psi \, = \, 0 \quad \Rightarrow \quad T_{(\,ij}\,\g_{\,k\,)}\, \cS \, = \, 0 \, . \label{eq_fermi}
\end{align}
The $\cB_{\,ijkl}$ and the $\cY_{\,ijk}$ are constructs combining the Lagrange multipliers $\b_{\,ijkl}$ and $\l_{\,ijk}$ with the fields. Barring some subtleties described in \cite{mixed_bose,mixed_fermi} and related to the linear dependence of the Labastida constraints, they can be regarded as gauge invariant completions of the Lagrange multipliers. Eliminating them in terms of $\cF$ or $\cS$ leads to the projected field equations of the constrained theory, so that one can directly refer to eqs.~\eqref{eq_bose} and \eqref{eq_fermi} in both settings. Furthermore, the field equations following from the Lagrangians \eqref{lag_unc} can be also cast in this form, via a partial gauge fixing that eliminates the compensators. In conclusion, one can focus on \eqref{eq_bose} and \eqref{eq_fermi} in order to analyze the consequences of the Lagrangian field equations that emerge in the various setups we described.

The main issue is then to understand whether \eqref{eq_bose} and \eqref{eq_fermi} are equivalent to the non-Lagrangian Labastida field equations \eqref{equations}. In the mixed-symmetry case the answer is rather subtle, since the Lagrangian field equations in general \emph{do not} reduce directly to the conditions
\be \label{eq_laba}
\left\{
\begin{array}{l}
\cF \, = \, 0 \, , \\
\cB_{\,ijkl} \, = \, 0 \, ,
\end{array}
\right.
\qquad\quad\qquad
\left\{
\begin{array}{l}
\cS \, = \, 0 \, , \\
\cY_{\,ijk} \, = \, 0 \, .
\end{array}
\right.
\ee
Indeed, non-trivial solutions of the homogeneous equations \eqref{eq_bose} and \eqref{eq_fermi} exist in some particular low space-time dimensions. However, in all these cases new gauge transformations of the fields and of the multipliers emerge. They can be used to gauge away the leftover quantities so that the reduction to the Labastida form \eqref{eq_laba} can be completed in this roundabout way. The Lagrangian field equations thus describe the propagation of the correct number of degrees of freedom, since the gauge fields satisfy \eqref{equations} and the multipliers are expressed in terms of them\footnote{In \cite{mixed_bose,mixed_fermi} the reduction of the field equations to the Labastida form is analyzed in detail for two-family fields. The result is that new symmetries involving the gauge fields only emerge in cases corresponding to trivial irreps of the little group $so(D-2)$. However, even if strictly speaking there is no dynamics in the pathological models, the reduction to the Labastida form is still crucial in order to prove that the $gl(D)$ fields actually do not propagate any degrees of freedom.}. The example of linearized gravity, that we often recalled here, can help to better qualify this phenomenon. In fact, in two dimensions the linearized Einstein tensor of \eqref{lag_grav} coincides with the traceless part of the linearized Ricci tensor, so that the Lagrangian field equation cannot provide any information on its trace. However, two-dimensional gravity is invariant under Weyl transformations, and their linearized version
\be \label{Weyl}
\d \, h_{\,\m\n} \, = \, \h_{\,\m\n} \, \O
\ee
suffices to set to zero the leftover trace. For symmetric fields only a gravitino in two dimensions presents a similar behavior, but in the mixed-symmetry case a rather rich set of models that are invariant under linearized Weyl-like symmetries exists \cite{mixed_bose,mixed_fermi}. Moreover, rewriting eq.~\eqref{lag_grav} in terms of the field $h_{\,\m\n}$ as in \eqref{antigrav} makes it manifest that the Lagrangian of two-dimensional gravity is a total derivative in $D=2$, where the same is true also for the Rarita-Schwinger Lagrangian \eqref{antirarita}. The rewriting \eqref{antilagb} of the Lagrangians \eqref{lag_bose} for two-column Bose fields and the rewriting \eqref{antilagf} of the Lagrangians \eqref{lag_fermi} for fully antisymmetric Fermi fields make it clear that these cases are only the first elements of a wider class of fields with a similar behavior. On the other hand, in the mixed-symmetry case the correspondence between Weyl-like symmetries and Lagrangians that are total derivatives is not one-to-one. The analysis performed in \cite{mixed_bose,mixed_fermi} for \emph{two-family} fields actually shows that in $D\leq4$ the models that do not propagate any degrees of freedom comprise three distinct classes: those without extra symmetries, those invariant under Weyl-like gauge transformations but with non-trivial Lagrangians and finally those with Lagrangians that are total derivatives. Although no degrees of freedom are involved in these pathological models, the rich structure of two-dimensional gravity suggests that they could well encode interesting properties.

%%%%%%%%%%%%%%%%%%%%%%%%%%%%%%%%%%%%%%%%%%%%%%%

\vspace{25pt}
\noindent
{\bf Acknowledgments}\hspace{5pt} I would like to thank D.~Francia, J.~Mourad and A.~Sagnotti for discussions and collaboration. I am also grateful to the APC-Paris VII, to the Chalmers University of Technology, to the \'Ecole Polyt\'echnique and to the ``Galileo Galilei Institute'' (GGI) of Florence for their kind hospitality.  This work was supported in part by Scuola Normale, by INFN, by the MIUR-PRIN contract 2007-5ATT78 and by the EU contracts MRTN-CT- 2004-503369 and MRTN-CT-2004-512194.

%%%%%%%%%%%%%%%%%%%%%%%%%%%%%%%%%%%%%%%%%%%%%%%%%%%%%%%%%%%%%%%%%%%%%

\bibliographystyle{amsplain}

\end{document}